# Построение Динамической Системы Опережающего Риск – Менеджмента и Оценки Рисков Компании

Денис С. Гусев gysev8500@mail.ru, Елена Г. Демидова, dmitrikeg@mail.ru, Ольга А. Новикова olga090984@yandex.ru

Старооскольский технологический институт им. А.А. УГАРОВА(филиал) федерального государственного автономного образовательного учреждения высшего образования«Национальный исследовательский технологический университет

**Abstract.** Целью исследования, представленного в данной статье, является разработка динамической системы прогнозирования и минимизации рисков промышленной компании на основе их количественной оценки. В статье рассмотрен понятийный аппарат сущностного содержания управления рисками промышленного предприятия, проведен обзор теоретических аспектов систем риск – менеджмента и наиболее значимых с практической точки зрения методов управления рисками. Расширен методологический аппарат качественного и количественного анализа и оценки рисков на основе выявленной некоторой условности классификационных признаков рисков и предложен системный подход к классификации рисков промышленного предприятия с учетом динамики их влияния на объект, приведены этапы построения динамической системы управления рисками. В статье обоснована необходимость дополнения динамической системы управления рисками промышленных предприятий методиками качественной и количественной оценки рисков с целью формирования эффективных стратегий риск – менеджмента и минимизации негативного воздействия рисков. Представленный диапазон количественной оценки рисков в зависимости от вероятности наступления позволяет систематизировать поведение рисков, а все многообразие поведения рисков сводится к восьми стратегиям их поведения. Научно обосновано, что наибольшим потенциалом негативного воздействия на управляемый объект обладают растущие существующие риски со средней или высокой вероятностью возникновения, за ними следуют возможные растущие риски с аналогичной вероятностью возникновения. Это позволяет установить критический количественный диапазон каждого риска и на этой основе осуществить количественную оценку рисков, а также оценить качество риск – менеджмента на основе расчета интегрального показателя рисков системы.

# 1 Introduction

Свободное экономическое хозяйствование определяет не только эффективную деятельность компании, но и служит источником постоянно изменяющихся внешних и внутренних факторов, которые в своей совокупности образуют динамическую взаимосвязанную и взаимозависимую систему. Причем влияние этих факторов может иметь как положительное, так отрицательное влияние на эффективность производственно – хозяйственной деятельности компании.

Снижение негативного влияния изменений и минимизация потерь от негативного воздействия факторов деловой среды реализуется на практике посредством риск – менеджмента.

Теоретические аспекты концептуальных основ риск-ориентированного подхода осуществления промышленно-хозяйственной деятельности широко представлены Трейманом М.Г. (Trejman M.G.&Varygina O.S.,2017), Латробе Дж., Барри Е. (Latrobe J. & Barry E.,2017), Валеевым С. (Valeev S. & Kondratyeva N., 2021), Панягиной А.Е. (Panyagina A.E., 2012) Картвелишвили В. М. (V. M. Kartvelishvili & O. A. Sviridova, 2017) и другими. По их мнению, управлять риском - значит минимизировать его или устранять, а система управления рисками базируется на исследование и анализе вероятности наступления тех или иных видов риск (Lankina S.A.&Flegontov V.I.,2015).

В работах Сенькова А.В. и Бобрякова А.В (Sen'kov A.V.&Bobryakov A.V., 2016) представлено унифицированное описание рисков, которое основывается на теоретико-множественном подходе и обеспечивает представление всех элементов и сущностей, участвующих в управлении рисками, однако, в современных экономических условиях промышленные компании функционируют в деловой среде с широким спектром постоянно меняющихся внешних и внутренних рисков, возникает необходимость не только постоянного мониторинга, анализа и оценки существующих рисков, но и разработки специфических инструментов риск - менеджмента, позволяющих осуществлять прогнозирование и опережающее управление возможными последствиями рисков, с целью нейтрализации или минимизации их негативного влияния на управляемую систему.

Для этого необходимо уточнить понятийный аппарат сущностного содержания понятия «управление рисками», дополнить существующую классификацию рисков промышленного предприятия дополнительными классификационными признаками с учетом динамики их влияния на объект, предложить систему опережающего риск – менеджмента промышленного предприятия на основе анализа динамики влияния рисков на управляемый объект и предложить методику количественной оценки рисков системы.

## 2 Materials and Methods

Изучение работ в области риск – менеджмента показало, что анализ риска предполагает оценку вероятности его наступления, возможных негативных последствий и их количественную оценку с использованием методик качественного и количественного анализа и оценки рисков.

Наиболее широко применяемые на практике динамические методы управления рисками представлены в таблице 1.

**Таблица 1.** Основные динамические методы управления рисками

| Методы | Методы сценарного подхода | Методы имитационного моделирования |
|---|---|---|
| Краткая характеристика | Построение сценариев по принципу оптимистический, пессимистический, реалистичный. Выбор критериев эффективности, сравнительный анализ показателей с базисными значениями, разработка корректирующих воздействий. | Описание изучаемого объекта и повторение его поведенческих характеристик на основе математической статистики |
| Эмпирическая база | Гипотезы экспертов, составленные о величине и влиянии конкретизированного риска. | На основе системного анализа, выбора шкалы риска и значений показателей измерения риска, определяется уровень риска. |
| Результат | Помогает выявить оценочные критерии, влияющие на основные показатели | Разработка матрицы последствий и вероятностей, сравнение степени и уровня риска с его заданными критериями. Градация рисков на незначительные, значительные и катастрофические |
| Преимущества | Простота | Простота в использовании |
| Недостатки | Субъективизм | Характеризует отдельные ситуации |

*Source:* (Compiled by the authors).

Таким образом, наибольшее практическое применение имеют методы сценарного подхода и имитационного моделирования, использование которых дает возможность выработать и оценить целесообразность мероприятий по минимизации рисков. Однако, данные методы имеют некоторые недостатки, влияющие на качество управленческих решений. Использование методов сценарного подхода возможно только при четком понимании вариантов развития событий, а методы имитационного моделирования характеризуют только отдельно взятые ситуации.

Однако, все риски воздействуют на объект в совокупности, непрерывно изменяясь и взаимодействуя между собой, что изменяет качество управляемого объекта. Это свидетельствует об условности классификационных признаков рисков, а сам риск является возможным событием с большей или меньшей степенью вероятности наступления и с разными последствиями при его наступлении.

Поэтому, на наш взгляд, управлять рисками, то есть вероятными событиями, невозможно, но возможно управлять экономической системой, так, что при наступлении риска его последствия будут минимальны.

Цель исследования заключается в разработке динамической системы прогнозирования и минимизации рисков промышленной компании на основе их количественной оценки.

Это диктует необходимость дополнить существующую классификацию рисков промышленного предприятия дополнительными классификационными признаками с учетом динамики их влияния на объект, а именно: по степени влияния и по динамике влияния на объект. Выделение этих классификационных признаков позволяет более точно идентифицировать риски конкретного предприятия и построить цепочку зависимости рисков, а также спрогнозировать поведение управляемой системы при изменении динамики и степени влияния определенного риска.

Проведенный анализ методов идентификации рисков и построение модели опережающего риск – менеджмента позволяет предложить методику количественной оценки рисков.

## 3 Results

Оценка рисков управляемой системы в рамках предложенной модели опережающего риск – менеджмента может быть осуществлена на основании следующих допущений.

Существование риска признается, если имеет место неравенство $Rj > 0$,
если $Rj$ меньше 0, то риск можно признать не существенным.

Поэтому, система оценки рисков может быть представлена как сумма влияния на объект возможных и существующих рисков:

$$R = R_в + R_с \quad (1),$$

Где $R_в$ – все выявленные возможные риски,

$R_с$ - все существующие риски, влияющие на объект управления.

Все выявленные риски $R_в$ принадлежат к множеству $V$ возможных снижающихся $r_{cv}$ и растущих $r_{rv}$ рисков с диапазоном количественной оценки от -∞ до 0,01.

Все риски $R_с$ принадлежат к множеству $S$ существующих снижающихся $r_{cs}$ и растущих $r_{rs}$ рисков с диапазоном количественной оценки от 0,01 до 0,99. То есть,

$R_в \in V\{r_{cv}; r_{rv}\};$

Rc∈S{$r_{cs}$; $r_{rs}$}.
При этом:
Rв={$r_v$∈V | $r_v$≤0,01};
Rc={$r_s$∈ S | 0,01≤$r_c$≤ 0,99}.
Область Fкатастрофических рисков $r_k$= [1;+∞).
Графическая интерпретация данных представлена на рисунке 1.

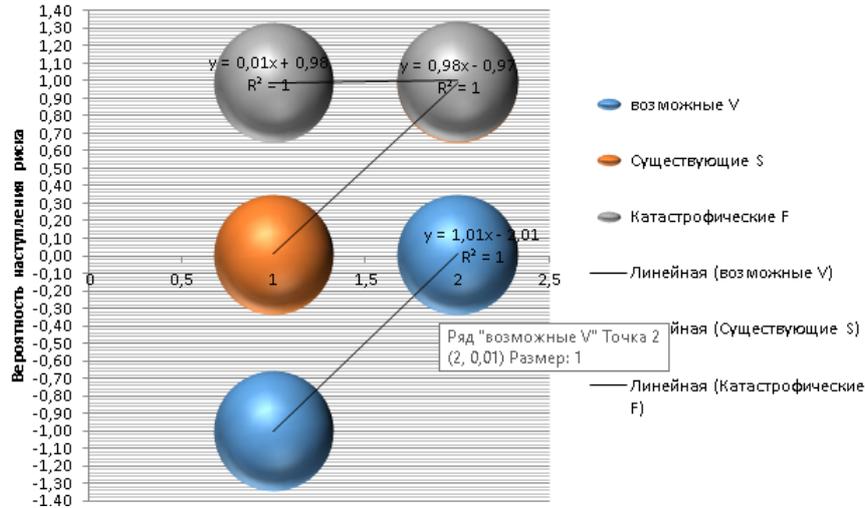

Рисунок 1. Прогнозирование поведения и оценка уровня риска на основании линии тренда по функции *x = f (y). Source:* (Compiled by the authors).

Анализ данных, представленных на рисунке позволяет заключить, что количественная оценка уровня риска будет сводится к вычислению функции x = f (y). Каждому значению области определения «x» данной функции соответствует свой уникальный «y», и наоборот – по любому значению «y» представляется возможным однозначно установить «x». Таким образом, это биективная функция. То есть на основе экспертно установленной вероятности наступления риска представляется возможным вычислить значение x, то есть дать количественную оценку риска.

К - критическое значение анализируемого риска по степени влияния на объект.

Диапазон количественной оценки рисков в зависимости от вероятности наступления позволяет систематизировать поведение рисков, а все многообразие поведения рисков проекта может быть сведено к восьми стратегиям поведения, представленным в таблице 2.

**Таблица 2.** Систематизация поведения рисков.

| Характеристика риска по мету | Вероятный / Существующий | Растущий↑ /Снижающийся↓ | Вероятность | Степень допустимости риска | Диапазон оценки по вероятности насту | Диапазон количественной оценки | Критическое значение |
|---|---|---|---|---|---|---|---|

| возникновения | | | | | пления риска | | ($x = f(y)$) |
|---|---|---|---|---|---|---|---|
| Внешний | Вероятный | ↓ | Низкая | Не существенный/Допустимый | от 0,99 до 0,01 | от 0 до 0,49 | 0,4851 |
| | Вероятный | ↑ | Средняя/Высокая | Допустимый/Критический | от 0,01 до 0,99 | от 0,5 до 1,99 | 1,99 |
| | Существующий | ↓ | Низкая | Допустимый | 1 | от 0,99 до 0,5 | 0,49 |
| | Существующий | ↑ | Средняя/высокая | Допустимый/Критический | 1 | от 0,5 до 0,99 | 0,99 |
| Внутренний | Вероятный | ↓ | Низкая/Средняя | Допустимый | от 0,99 до 0,01 | от 0 до 0,49 | 0,4851 |
| | Вероятный | ↑ | Средняя/Высокая | Допустимый/Критический | от 0,01 до 0,99 | от 0,5 до 1,99 | 1,99 |
| | Существующий | ↓ | Низкая/Средняя | Допустимый | 1 | от 0,99 до 0,5 | 0,49 |
| | Существующий | ↑ | Средняя/Высокая | Допустимый/Критический | 1 | от 0,5 до 0,99 | 0,98 |

*Source:* (Compiled by the authors).

Для выявленных возможных рисков с целью их количественной оценки и глубины их влияния на объект управления установлены критические значения. Риск со средней вероятностью наступления, достигший критического значения переходит в разряд риска с высокой вероятностью наступления, риск с высокой вероятностью становится критическим при достижении значения 1, а вся взаимосвязанная система рисков тем более уязвима, чем ближе интегральный показатель рисков к 1. Катастрофическим риском системы может стать любой существующий растущий риск, достигший значения 1 или вероятный растущий риск, достигший значения 1 и трансформировавшийся в существующий.

Установив вероятность наступления риска представляется возможным рассчитать уровень его влияния на объект то есть x = f (y), общий уровень уязвимости управляемой системы может быть представлен произведением уровней каждого риска, большего 0, то есть существенного:

$$E_p = x_1 * x_2 * x_3 * \ldots x_n \qquad (2)$$

$E_p$ – интегральный показатель риска управляемой системы, значение которого позывает общий уровень риска всей системы. При чем, чем ближе интегральный показатель к 1, то есть 100%, тем более уязвима вся система.

В итоге можно заключить, что наибольшим потенциалом негативного воздействия на управляемый объект обладают растущие существующие риски со средней или высокой вероятностью возникновения, за ними следуют возможные растущие риски с аналогичной вероятностью возникновения. Поэтому система опережающего риск - менеджмента в первую очередь должна быть нацелена именно на растущие возможные или существующие риски со средней или высокой вероятностью возникновения, сосредоточив на их минимизации основные векторы управляющего воздействия.

## 4   Discussion

Дополнение классификации рисков по степени влияния на объект как существующие или вероятные, а по динамике влияния на объект как растущие или снижающиеся позволяет осуществить построение системы риск - менеджмента в определенной последовательности, состоящей из девяти этапов (Demidova E.G., Gusev D.S., Novikova O.A, 2020).

Так, первый этап системы опережающего риск – менеджмента предполагает выбор этапа реализации проекта от напрямую зависит временной горизонт выстраиваемой системы. Далее риски сгруппируются по уровням экономической системы, то есть сферам их возникновения (Samarina, V., Skufina, T., Samarin, A., Baranov, S., 2016). На третьем этапе необходимо выявить конкретные виды рисков (Samarina V., Skufina T., Samarin A., Ushakov D., 2019). На четвертом и пятом этапах осуществляется подготовка качественной базы для оценки рисков на анализа трендов развития экономической системы для каждого анализируемого уровня, осуществляется оценка величины и вектора воздействия каждого выявленного риска. Реализация представленных этапов опережающей системы риск - менеджмента позволяет идентифицировать риски, поэтому они объединены в первый блок. Мероприятия второго блока системы позволяют установить вероятность возникновения риска и, для каждого существующего риска, устанавливается степень допустимости. При этом внешние критические и катастрофические риски требуют пересмотра стратегических целей, так как их наступление неизбежно повлечет за собой возникновение критических и катастрофических внутренних рисков системы (Skufina T.P., Baranov S.V., Samarina V.P., 2016).

Далее необходимо выявить динамику влияния каждого вида риска, установить их взаимозависимость и дать качественную оценку (Langdalen H., Abrahamsen E. B., Selvik J. T., 2020). На основании такого анализа разрабатывается комплекс мероприятий по минимизации финансовых потерь именно от существующих растущих рисков

Реализация представленных этапов построения системы опережающего риск - менеджмента дает возможность идентифицировать, обобщить и систематизировать поведение рисков по таким их основным характеристикам как: внешние и внутренние, снижающиеся и растущие, вероятные и существующие, что позволяет в дальнейшем установить уровень допустимости риска и разработать мероприятия по минимизации их негативных последствий.
.

## 5   Conclusion

Таким образом, дополнение существующей классификации рисков такими классификационными признаками как: по степени влияния и по динамике влияния на объект, позволяет более точно идентифицировать риски конкретного предприятия и построить цепочку взаимозависимости рисков, а также спрогнозировать поведение управляемой системы при изменении динамики и степени влияния определенного риска. Представленный диапазон количественной оценки рисков в зависимости от вероятности наступления позволяет систематизировать поведение рисков, а все многообразие поведения рисков сводится к восьми стратегиям их поведения. Научно обосновано, что наибольшим потенциалом негативного воздействия на управляемый объект обладают растущие существующие риски со средней или высокой вероятностью возникновения, за ними следуют возможные растущие риски с аналогичной вероятностью возникновения.

В основе предложенной методики количественной оценки рисков лежит предположение о том, что на управляемую систему оказывают влияние как возможные, так и уже существующие риски. При этом поведение рисков управляемой системы характеризуется с позиции их динамики, то есть роста или снижения. Поэтому, на основе экспертно установленной вероятности наступления риска представляется возможным вычислить значение функции $x = f(y)$, то есть дать количественную оценку риска и рассчитать интегральный показатель риска управляемой системы, значение которого позывает общий уровень риска всей системы. Теоретически обосновано, что чем ближе интегральный показатель к 1, тем более уязвима вся система.

Практическая значимость предложенной методики идентификации и количественной оценки рисков состоит в том, что позволяет выявить наиболее значимые растущие возможные или существующие риски со средней или высокой вероятностью возникновения, сосредоточив на их минимизации основные векторы управляющего воздействия. Расчет интегрального показателя рисков системы позволяет оценить эффективность

риск – менеджмента. Дальнейший научный интерес представляет разработка взаимосвязанной системы влияния рисков на конкретные показатели эффективности деятельности компании и разработка алгоритма действий в значимости от поведения рисков.